\documentstyle[12pt]{article}
  \newlength{\extralineskip}
\newcommand{\beq}{\begin{equation}}
\newcommand{\eeq}{\end{equation}}
\newcommand{\bd}{\begin{displaymath}}
\newcommand{\ed}{\end{displaymath}}

\font\twlmsy=msbm10 at 12pt
\font\sevenmsy=msbm8
\font\fivemsy=msbm6
\newfam\Bbbfam
\textfont\Bbbfam=\twlmsy
\scriptfont\Bbbfam=\sevenmsy
\scriptscriptfont\Bbbfam=\fivemsy
\def\Bbb{\fam\Bbbfam}

\def\e{\, {\rm e}}

\def\tr{{\rm tr}}
\makeatletter
\newdimen\normalarrayskip              
\newdimen\minarrayskip                 
\normalarrayskip\baselineskip
\minarrayskip\jot
\newif\ifold             \oldtrue            \def\new{\oldfalse}
\def\arraymode{\ifold\relax\else\displaystyle\fi} 
\def\@arrayskip{\ifold\baselineskip\z@\lineskip\z@
     \else
     \baselineskip\minarrayskip\lineskip2\minarrayskip\fi}
\def\@arrayclassz{\ifcase \@lastchclass \@acolampacol \or
\@ampacol \or \or \or \@addamp \or
   \@acolampacol \or \@firstampfalse \@acol \fi
\edef\@preamble{\@preamble
  \ifcase \@chnum
     \hfil$\relax\arraymode\@sharp$\hfil
     \or $\relax\arraymode\@sharp$\hfil
     \or \hfil$\relax\arraymode\@sharp$\fi}}
\def\@array[#1]#2{\setbox\@arstrutbox=\hbox{\vrule
     height\arraystretch \ht\strutbox
     depth\arraystretch \dp\strutbox
     width\z@}\@mkpream{#2}\edef\@preamble{\halign \noexpand\@halignto
\bgroup \tabskip\z@ \@arstrut \@preamble \tabskip\z@ \cr}%
\let\@startpbox\@@startpbox \let\@endpbox\@@endpbox
  \if #1t\vtop \else \if#1b\vbox \else \vcenter \fi\fi
  \bgroup \let\par\relax
  \let\@sharp##\let\protect\relax
  \@arrayskip\@preamble}
\makeatother

\begin{document}

\begin{titlepage}

\baselineskip=12pt

\rightline{OUTP-96-74P}
\rightline{hep-th/9612251}
\rightline{   }
\rightline{December, 1996}

\vskip 0.6truein
\begin{center}

\baselineskip=24pt
{\Large\bf D-brane Configurations and Nicolai Map in Supersymmetric Yang-Mills
Theory}\\
\baselineskip=12pt
\vskip 0.8truein
{\bf I. I. Kogan}$^a$, {\bf G. W. Semenoff} $^{b,}$\footnote{\baselineskip=12pt
Work supported in part by the Natural Sciences and Engineering Research Council
of Canada.} and {\bf R. J. Szabo}$^{a,1}$\\

\vskip 0.3truein

$^a$ {\it Department of Theoretical Physics\\ University of Oxford\\ 1 Keble
Road, Oxford OX1 3NP, U.K.}\\

\bigskip

$^b$ {\it Department of Physics and Astronomy\\ University of British
Columbia\\ Vancouver, B.C. Canada V6T 1Z1}\\

\vskip 1.3 truein

\end{center}

\begin{abstract}

\baselineskip=12pt

We discuss some properties of a supersymmetric matrix model that is the
dimensional reduction of supersymmetric Yang-Mills theory in ten dimensions and
which has been recently argued to represent the short-distance structure of M
theory in the infinite momentum frame. We describe a reduced version of the
matrix quantum mechanics and derive the Nicolai map of the simplified
supersymmetric matrix model. We use this to argue that there are no phase
transitions in the large-$N$ limit, and hence that S-duality is preserved in
the full eleven dimensional theory.

\end{abstract}

\end{titlepage}

\clearpage\newpage

\baselineskip=18pt

The conventional understanding of the spacetime structure of string theory has
drastically changed over the last few years. It has been realized recently that
all ten dimensional superstring theories are related by non-perturbative
dualities and that they can be thought of as originating, via Kaluza-Klein
types of compactifications, from a single, eleven dimensional theory known as
`M theory' (see \cite{duff} for recent reviews). The dynamics of this theory
are
not yet fully understood. Some of the central objects in understanding the
string dualities are non-perturbative, $p$-dimensional degrees of freedom known
as D$p$-branes \cite{polch}, on which the endpoints of strings can attach (with
Dirichlet boundary conditions). The low-energy dynamics of a system of $N$
parallel D-branes can be described by an $N\times N$ matrix model obtained from
the dimensional reduction of ten dimensional supersymmetric Yang-Mills theory
with gauge group $U(N)$ \cite{dewit,wittenD}. The large-$N$ limit of this
matrix model has been recently conjectured to describe the small distance
spacetime structure of M theory in the infinite momentum frame \cite{bfss}. The
explicit solution of this matrix model therefore has the potential of providing
a non-perturbative description of the largely unknown dynamical objects
describing the short distance behaviour of the full eleven dimensional theory
\cite{town}.

In this Letter we will discuss some basic properties of the supersymmetric
matrix model introduced in \cite{wittenD,bfss}. We examine a particular
reduction of the model proposed in \cite{wittenD,bfss} to static D-brane
configurations with transverse $SO(8)$ rotational symmetry. We explicitly
construct the Nicolai map associated with the supersymmetry in this reduced
model and use it to analyse Schwinger-Dyson equations of the matrix model. We
show that the results from this analysis are consistent with other known
results of the D-brane field theory, and also that this simple approach gives
some insights into the structure of the full matrix model. In particular, the
reduced model seems to have no phase transitions in the large-$N$ limit and
S-duality is preserved in this representation of the full eleven-dimensional
theory.

First, we discuss some aspects of the representation of systems of D-branes by
Yang-Mills fields. Consider the gauged supersymmetric matrix quantum mechanics
with action \cite{wittenD,bfss}
\beq
S=\frac1{2g}\int
dt~\tr\left(\sum_{i=1}^9\left(D_tX^i\right)^2-\sum_{i<j}\left[X^i,X^j\right]^2
+2\psi^\alpha D_t\psi_\alpha-2\sum_{i=1}^9\psi^\alpha(\gamma_i)_\alpha^\beta
\left[\psi_\beta,X^i\right]\right)
\label{qmaction}\eeq
where $D_tY=\partial_tY-i[A_0(t),Y]$ is the temporal component of the gauge
covariant derivative, the trace is taken over unitary group indices, and we
have chosen units in which the string tension is $\alpha'=1/2\pi$.
Here $X^i(t)=[X^i_{ab}(t)]$, $a,b=1,\dots,N$, $i=1,\dots,9$, are $N\times N$
Hermitian matrices in the adjoint representation of $U(N)$ which are obtained
as the spatial components in the reduction to (0 + 1) dimensions of a (9 + 1)
dimensional $U(N)$ Yang-Mills field $A_\mu(x,t)$, $\mu=0,\dots,9$. They
describe the collective coordinates of a system of $N$ parallel
D0-branes (with infinitesimal separation), and they transform under the vector
representation of the rotation group $SO(9)$ of the space transverse to the
compactified eleventh dimension of the underlying supergravity theory. The
superpartners of the matrices $X^i$ are the Majorana spinors
$\psi^\alpha(t)=[\psi^{\alpha}_{ab}(t)]$, $\alpha=1,\dots,16$, which transform
under the 16-dimensional spinor representation of $SO(9)$, and under the
adjoint representation of the gauge group $U(N)$. The Dirac matrices $\gamma_i$
are the generators of the spin(9) Clifford algebra
\beq
\left\{\gamma_i,\gamma_j\right\}=2\delta_{ij}
\label{cliffalg}\eeq
in a Majorana basis. The coupling constant $g$ is related to the eleven
dimensional compactification radius $R$ by
\beq
R=g^{2/3}l_P
\label{comprad}\eeq
where $l_P$ is the eleven dimensional Planck length.

The action (\ref{qmaction}) describes the short-distance properties of
D0-branes in weakly-coupled type-IIA superstring theory \cite{wittenD,town}. It
was argued in \cite{bfss} to be the most general infinite momentum frame action
with at most two derivatives which is invariant under the $U(N)$ gauge group
and the full eleven dimensional Lorentz group. It is further invariant under
the
infinitesimal ${\cal N}=1$ supersymmetry transformation
\beq\new{\begin{array}{l}
\delta_\varepsilon X^i=-2\varepsilon^\alpha(\gamma^i)_\alpha^\beta\psi_\beta
\\\delta_\varepsilon\psi^\alpha=\frac12\left(\sum_iD_tX^i(\gamma_i)^\alpha
_\beta+\frac12\sum_{i<j}\left[X^i,X^j\right]\left[\gamma_i,\gamma_j\right]^
\alpha_\beta\right)\varepsilon^\beta\\\delta_\varepsilon
A_0=-2\varepsilon^\alpha\psi_\alpha\end{array}}
\label{susyqmtransf}\eeq
where $\varepsilon^\alpha$ are 16 global Majorana spinor parameters. There
are another set of 16 supersymmetries which are realized trivially as
$\delta_{\varepsilon'}\psi^\alpha=\varepsilon'^\alpha$,
$\delta_{\varepsilon'}X^i=\delta_{\varepsilon'}A_0=0$, where $\varepsilon$ and
$\varepsilon'$ are independent supersymmetry parameters. Together, these two
sets of supersymmetry transformations (with slight modifications) generate the
full 32-dimensional ${\bf16}\oplus{\bf16}$ representation of the super-Galilean
group of the eleven dimensional theory in the light-cone frame. A number of
properties of M theory have been verified using the action (\ref{qmaction})
\cite{bfss},\cite{berk}--\cite{doug}.

It may seem puzzling at first sight that a dimensionally-reduced gauge field
$A$ yields the appropriate D-brane coordinatization. What is intriguing though
is that string theoretic T-duality is a key to this connection. If we
compactify the first spatial dimension onto a circle $S^1_{\cal R}$ of radius
$\cal R$ (or more generally several dimensions onto a torus), then the
angular coordinate $x^1$ (or coordinates $x^i$) takes values in the interval
$x^1\in[0,2\pi{\cal R}]$. The original supersymmetric Yang-Mills theory in ten
dimensions describes the low-energy sector of open superstrings, and in this
theory there exists other gauge-invariant variables, namely the path-ordered
Wilson loop operators
\beq
W[A]=~\tr~P\exp\left(i\oint_{S^1_{\cal R}}A_i~dx^i \right)
\label{wilsons}\eeq
which are invariant under the large gauge transformations which wind around the
compactified direction. The connection appearing in the argument of the
exponential in (\ref{wilsons}) lies in the adjoint representation of $U(N)$.
The corresponding classical gauge field orbits are
topological and lie in the interval $A_1\in[0,\frac{2\pi}{\cal R}]$. For each
coordinate $x^i$ living on a circle $S^1_{{\cal R}_i}$ there is a ``dual''
coordinate $X^i = \alpha' A_i$ which also lives on a circle $S^1_{r_i}$ but
with a dual radius $r_i=\alpha'/{\cal R}_i$. As we have mentioned the
ten-dimensional supersymmetric Yang-Mills theory describes the low-energy
dynamics of open strings with Neumann boundary conditions. Under T-duality
${\cal R}\to\alpha'/{\cal R}$ the Neumann boundary conditions are transcribed
into Dirichlet boundary conditions. But this means that the topological gluon
degrees of freedom are converted into the D-brane fields $X(t)=\alpha'A(t)$
describing open strings with Dirichlet boundary conditions. Thus the
ten-dimensional gluon fields describe the dual theory of the D-branes and
T-duality naturally identifies the topological orbits of the Yang-Mills fields
as the D-brane coordinates. The role of T-duality in the matrix model
(\ref{qmaction}) has also been addressed from other points of view in
\cite{Tduality}.

A question which now arises is how to really measure the original angular
coordinate $x^i$ given the gauge field coordinate $X^{i}$. The solution is to
consider more complicated objects, such as the abelian Wilson
loops with a winding number $n\in{\Bbb Z}$
\beq
W_n [A]=\exp\left(i n\oint_{S^1_{\cal R}}A_i~dx^i \right)
\label{wilsons1}\eeq
which are associated with a single D-brane configuration. In the general case,
i.e. in the case of a non-abelian gauge group, we should consider Wilson loops
in different representations of $U(N)$, but for the sake of illustration we
shall discuss here only the $U(1)$ Wilson loops (\ref{wilsons1}). This
simplication can be thought of as a compactification of all of the ten
dimensions on which the group variables (\ref{wilsons}) are restricted to the
maximal torus of the $U(N)$ gauge group. Using these objects and T-duality one
can in principle obtain the angular coordinate $x^i$ through a superposition
$\sum_{n\in{\Bbb Z}}W_n[A]$ of  different Wilson loops (using harmonic analysis
on unitary groups in the general case).

In fact, this construction demonstrates
that the Wilson loops in the ``N-theory" (i.e. ordinary open strings with
Neumann boundary conditions) are equivalent to the vertex operators in the
``D-theory" (open strings with Dirichlet boundary conditions). To see this, we
expand the topological gauge field configurations of the low-energy description
of strings on the compactified space in Fourier modes as
\beq
A_1(t,x)=\sum_{\ell\in{\Bbb Z}}A_1^{(\ell)}(t)\e^{i\ell
{\cal R}x}~~~~~,~~~~~x\in[0,2\pi/{\cal R}]
\label{1+1A}\eeq
These dual fields are described by a (1 + 1) dimensional gauge theory that is
also dimensionally reduced from the ten-dimensional supersymmetric Yang-Mills
theory \cite{Tduality}. Then the Wilson loop (\ref{wilsons1}) becomes
\beq
W_n=\e^{in{\cal R}A_1^{(0)}(t)}=\e^{ip_nX^1(t)}
\label{wilsonvert}\eeq
where $p_n=n/r=n{\cal R}/\alpha'$ is the momentum of the string winding mode in
the compactified direction. Thus T-duality converts the non-commuting position
matrices $X^i$ describing the D-brane configurations into gauge fields in the
dual theory, and also the winding number of the topological gauge field modes
into the string momentum in the compactified direction. In the general case
then, we can conjecture the equivalence between the string scattering
amplitudes of the D-theory defined by correlators of the vertex operators and
expectation values of the Wilson loop operators in the dual
N-theory\footnote{\baselineskip=12pt See \cite{ishi} for another discussion of
the relationship between string vertex operators and Wilson lines.},
\beq
\left\langle W_{\vec n_1}[A]\cdots W_{\vec
n_k}[A]\right\rangle_{N}=\left\langle\e^{i\vec{n}_1\cdot\vec{X}}\cdots\e^{i
\vec{n}_k\cdot\vec{X}}\right\rangle_D
\label{wilsons2}\eeq
where $\vec n_j\in{\Bbb Z}^9$ and we have defined $W_{\vec
n_j}[A]=\prod_{i=1}^9W_{(\vec n_j)^i}[A]$ in terms of the abelian Wilson loops
(\ref{wilsons1}). In the non-abelian case, the correlators (\ref{wilsons2})
will generalize in the appropriate way in terms of representations of $U(N)$.

We now examine the problem of obtaining an explicit solution of the matrix
model (\ref{qmaction}). For this, we further dimensionally reduce the theory
described by (\ref{qmaction}) to a zero-dimensional $N\times N$ supersymmetric
matrix model, i.e. we ignore the time dependence in (\ref{qmaction}) and work
in the Weyl gauge $A_0=0$. This means that we are studying the model
separately over each constant time slice describing a static configuration of
the D0-branes. This reduction can be thought of as originating by compactifying
the time direction of the ten dimensional Yang-Mills theory where the adjoint
representation fermions have periodic boundary conditions, and then taking the
limit in which the radius of
compactification vanishes. This simplification has the advantage of eliminating
non-local operators that would appear from the time-dependence. We shall
discuss the inclusion of time-dependent fields at the end of this Letter.

With this further reduction, the partition function of the model is given
by the finite-dimensional matrix integral
\beq
Z=\int\prod_{a,b}\prod_idX_{ab}^i~\prod_\alpha d\psi^\alpha_{ab}~\exp\left\{
\frac
N{2g}~\tr\left(\sum_{i<j}\left[X^i,X^j\right]^2-2\sum_i\psi^\alpha(\gamma_i)
_\alpha^\beta\left[\psi_\beta,X^i\right]\right)\right\}
\label{part0D}\eeq
We can expand the matrix integration variables in (\ref{part0D}) in a basis
$T^A$ of the unitary group as $X^i=X_A^iT^A$ and $\psi^\alpha=\psi_A^\alpha
T^A$, where $A=1,\dots,N^2$ and the Hermitian $U(N)$ generators satisfy
\beq
\left[T^A,T^B\right]=if^{AB}_{~~~C}T^C~~~~~,~~~~~\tr~T^AT^B=\frac12\delta^{AB}
\label{gennorm}\eeq
The integration over the Majorana fermions in (\ref{part0D}) is Gaussian
and can be evaluated explicitly using the Berezin integration rules for
$\psi_A^\alpha$. It produces a square root of the determinant determined by the
representation of the adjoint action of $X^i$, and (\ref{part0D}) becomes
\beq
Z=c_N\int\prod_i\prod_{D=1}^{N^2}dX_D^i~{\rm Pfaff}\left[\frac
i{2g}f^{ABC}\sum_i(\gamma_i)_\alpha^\beta X_C^i\right]\exp\left\{\frac
N{2g}\sum_{i<j}\tr\left[X^i,X^j\right]^2\right\}
\label{partfermint}\eeq
where the Pfaffian is taken over both the adjoint $U(N)$ representation indices
$A,B=1,\dots,N^2$ and the spin(9) indices $\alpha,\beta=1,\dots,16$. Here and
in the following $c_N$ denotes an irrelevant numerical constant.

We would now like to exploit the supersymmetry (\ref{susyqmtransf}) of the zero
dimensional model to compute correlation functions of the matrix model. When
the number of bosonic and fermionic degrees of freedom are the same, the
supersymmetry is maximal, in that it
holds even when the fields are off-shell. In the present model, the number of
fermionic and bosonic degrees of freedom do not match. Normally, supersymmetry
would require
on-shell fields and the addition of auxilliary fields to make the number of
physical boson and fermion modes equal. However, we can adjust things to match
by exploiting the original interpretation of the matrices $X^i$ from the
dimensional reduction of ten-dimensional supersymmetric Yang-Mills theory. The
latter theory can be gauge-fixed and quantized in the light-cone gauge, after
which there are only eight propagating gluon degrees of freedom corresponding
to the various possible transverse polarizations. Since a Majorana-Weyl spinor
in ten dimensions has eight physical modes, the minimal Yang-Mills action in
ten dimensions is supersymmetric without the need of introducing auxilliary
fields. Thus we match the collective degrees of freedom $X^i$ of the
system of D-branes with the physical modes of the full Yang-Mills theory by
reducing the target space degrees of freedom from nine to eight by setting
$X^9=0$ and working in the Majorana-Weyl representation of spin(9) in the
matrix model above. The constraint $X^9=0$ can be thought of as
a light-cone gauge fixing condition in the nine-dimensional transverse
space.
Although not precise from the point of view of the M theory dynamics, this
simplification produces a toy model that will shed
light on some of the properties of the nine-dimensional theory that we started
with\footnote{\baselineskip=12pt Some different reductions of the matrix
quantum mechanics (\ref{qmaction}) have also been suggested. In \cite{peri} it
was argued that the model can be truncated to zero dimensions by augmenting the
transverse rotational symmetry to $SO(11)$ and viewing the matrix model as the
dimensional reduction of supersymmetric Yang-Mills theory in (10 + 2)
dimensions. In \cite{ishi} it was argued that the D-brane field theory
associated with weakly-coupled type-IIB superstrings could be
viewed as the large-$N$ reduction of the ten dimensional supersymmetric
Yang-Mills theory.}.

With this simplification we now exploit some features of the group theory for
$SO(9)$. The Dirac generators of spin(9) in the Majorana-Weyl basis can be
constructed from the reducible ${\bf8}_s\oplus{\bf8}_c$ chiral representation
of spin(8) by decomposing the 16-dimensional gamma-matrices in the $8\times8$
block form
\beq
\gamma_i=\pmatrix{0&(\gamma_i)_\alpha^{\dot\alpha}\cr(\gamma_i)_{\dot\beta}^
\beta&0\cr}~~~~~,~~~~~i=1,\dots,8
\label{gammablock}\eeq
where $(\gamma_i)_{\dot\alpha}^\alpha=(\gamma_i^T)_\alpha^{\dot\alpha}$,
$\alpha,\dot\alpha=1,\dots,8$, are the Dirac generators of spin(8). The
Clifford
algebra (\ref{cliffalg}) is then equivalent to the equations
\beq
(\gamma_i)_\alpha^{\dot\alpha}(\gamma_j)^\beta_{\dot\alpha}+
(\gamma_j)_\alpha^{\dot\alpha}(\gamma_i)^\beta_{\dot\alpha}=
2\delta_{ij}\delta_\alpha^\beta
\label{spin8cliffalg}\eeq
and similarly with dotted and undotted chiral indices interchanged. The spin(8)
Dirac generators can be expressed explicitly as direct products of $2\times2$
block matrices by
\beq\new{\begin{array}{lll}
(\gamma_1)_\alpha^{\dot\alpha}=-i\sigma_2\otimes\sigma_2\otimes\sigma_2
{}~~~~~&,&~~~~~
(\gamma_2)_\alpha^{\dot\alpha}=i{\bf1}\otimes\sigma_1\otimes\sigma_2\\
(\gamma_3)_\alpha^{\dot\alpha}=i{\bf1}\otimes\sigma_3\otimes\sigma_2
{}~~~~~&,&~~~~~
(\gamma_4)_\alpha^{\dot\alpha}=i\sigma_1\otimes\sigma_2\otimes{\bf1}\\
(\gamma_5)_\alpha^{\dot\alpha}=i\sigma_3\otimes\sigma_2
\otimes{\bf1}~~~~~&,&~~~~~
(\gamma_6)_\alpha^{\dot\alpha}=i\sigma_2\otimes{\bf1}\otimes\sigma_1\\
(\gamma_7)_\alpha^{\dot\alpha}=i\sigma_2
\otimes{\bf1}\otimes\sigma_3~~~~~&,&~~~~~
(\gamma_8)_\alpha^{\dot\alpha}={\bf1}\otimes{\bf1}\otimes{\bf1}\end{array}}
\label{spin8expl}\eeq
where $\sigma_1=\pmatrix{0&1\cr1&0\cr}$, $\sigma_2=\pmatrix{0&i\cr-i&0\cr}$ and
$\sigma_3=\pmatrix{1&0\cr0&-1\cr}$ are the Pauli spin matrices. The remaining
spin(9) Dirac matrix is then
$\gamma_9=\gamma_1\gamma_2\cdots\gamma_8=\pmatrix{{\bf1}&0\cr0&-{\bf1}\cr}$.

The block decomposition (\ref{gammablock}) of the first eight gamma-matrices
shows that the Pfaffian in (\ref{partfermint}) with $X^9=0$ becomes squared,
with the spinor part of the determinant restricted to the spin(8) chiral
indices. Then we can write the partition function as
\beq
Z=c_N\int\prod_{i=1}^8\prod_{D=1}^{N^2}
dX^i_D~\det_{A,B;1\leq\alpha,\dot\alpha\leq8}
\left[\frac1{2g}f^{ABC}\sum_{i=1}^8(\gamma_i)^{\dot\alpha}_\alpha
X_C^i\right]\exp\left\{\frac N{2g}\sum_{1\leq
i<j\leq8}\tr\left[X^i,X^j\right]^2\right\}
\label{partchiral}\eeq
This effective reduction to eight dimensions matches bosonic and fermionic
degrees of freedom and allows us to exploit the supersymmetry in a simple way
to completely solve the matrix model. Essentially it enables us to use the
triality property of the eight-dimensional rotation group, i.e. that there
exists automorphisms between the vector and chiral spinor representations of
$SO(8)$.

We now label the eight spatial indices $i$ as the chiral indices
$\alpha,\dot\alpha$ of the spinor representation, and the first eight
components of the 16-component spinor field $\psi$ as the chiral parts
$\psi^\alpha$ and the last eight components as the anti-chiral parts
$\psi^{\dot\alpha}$ in the ${\bf8}_s\oplus{\bf8}_c$ decomposition of spin(9)
above. The static reduction of the quantum mechanical action
(\ref{qmaction}) can then be written in the standard form of an ${\cal N}=1$
supersymmetric field theory (after integration over superspace coordinates) as
\beq
S_0=-\frac1{2g}\left[\sum_{i,A}\left(\frac{\partial F}{\partial
X_A^i}\right)^2+\psi_A^\alpha\left(\frac{\partial^2F}{\partial
X_A^\alpha\partial
X_B^{\dot\alpha}}\right)\psi_B^{\dot\alpha}+\psi_A^{\dot\alpha}\left(\frac{
\partial^2F}{\partial X_A^{\dot\alpha}\partial
X_B^\alpha}\right)\psi_B^\alpha\right]
\label{S0super}\eeq
where the super-potential is
\beq
F(X)\equiv\frac13(\gamma_k)^{\dot\alpha}_\alpha~{\rm
tr}~X^k\left[X^\alpha,X_{\dot\alpha}\right]=\frac16
(\gamma_k)^{\dot\alpha}_\alpha f^{AB}_{~~~C}X_A^kX_B^\alpha X_{\dot\alpha}^C
\label{Fdef}\eeq
The representation (\ref{S0super}) can be derived using the symmetry properties
of the gamma-matrices (\ref{spin8expl}), the $U(N)$ Jacobi identity
\beq
f^{ABC}f^{ADE}=f^{ADB}f^{ACE}-f^{ADC}f^{ABE}
\label{jacobiid}\eeq
and the $SO(8)$ Fierz identity
\beq
(\gamma_i)^\alpha_{\dot\alpha}(\gamma_j)^{\dot\beta}_\alpha=\delta_{ij}
\delta_{\dot\alpha}^{\dot\beta}+(\gamma_{ij})_{\dot\alpha}^{\dot\beta}
\label{fierzid}\eeq
where we have introduced the spinor matrix
\beq
(\gamma_{ij})_\alpha^\beta=\frac12\left((\gamma_i)_\alpha^{\dot\alpha}
(\gamma_j)_{\dot\alpha}^\beta-(\gamma_j)_\alpha^{\dot\alpha}(\gamma_i)
_{\dot\alpha}^\beta\right)
\label{gammaij}\eeq
and similarly for $(\gamma_{ij})_{\dot\alpha}^{\dot\beta}$.

The form of the action (\ref{S0super}) identifies the Nicolai map (i.e. the
Hubbard-Stratonvich transformation for the bosonic potential
$\sum_{i<j}\tr[X^i,X^j]^2$) \cite{nicolai} of this
supersymmetric field theory as
\beq
W^A_k(X)\equiv\frac{\partial F}{\partial
X_A^k}=\frac12(\gamma_k)_\alpha^{\dot\alpha}f^{AB}_{~~~C}X_B^\alpha
X_{\dot\alpha}^C~~~~~{\rm
or}~~~~~W_k^{ab}=\frac12(\gamma_k)^{\dot\alpha}_\alpha[X^\alpha,
X_{\dot\alpha}]^{ab}
\label{nicolaimapdef}\eeq
{}From (\ref{S0super}) we see that the Jacobian factor $|\det[\partial
W_k^A/\partial X_B^j]|^{-1}$ which arises in the change of variables $X\to
W(X)$ in the partition function (\ref{partchiral}) will cancel exactly with the
determinant that comes from integrating out the chiral fermion fields. The
partition function is thus trivially a Gaussian Hermitian matrix integral and
is formally unity,
\beq
Z=\frac{c_N}{(2g)^{64N^2}}\int\prod_{i=1}^8\prod_{A=1}^{N^2}dW^A_i~\e^{-\frac
N{4g}(W_i^A)^2}=c_N
\label{Ztrivial}\eeq
The free energy $\log Z$ is thus trivially an analytic function of the coupling
constant $g$ everywhere and it does not exhibit any phase transitions, even in
the large-$N$ limit. Furthermore, the correlation functions which are invariant
under the supersymmetry transformations (\ref{susyqmtransf}) can be obtained by
differentiating the free energy with respect to the coupling constants of the
model (in an appropriate superspace formulation). Thus any supersymmetric
correlator of the model vanishes, which is just the standard
non-renormalization that usually occurs in supersymmetric field theories. The
existence of the Nicolai map and these implied properties of the matrix model
are essentially the content of the supersymmetric Ward identities.

The only observables of the matrix model which are non-trivial are those which
are not supersymmetric. To examine such correlation functions, we use the
Nicolai map (\ref{nicolaimapdef}) to express correlators $\langle\cdot\rangle$
of the original matrix model (normalized so that $Z=1$) as free Gaussian
averages $\langle\!\langle\cdot\rangle\!\rangle$ of the Nicolai field. For
instance, from $\langle\!\langle W_i^{ab}\rangle\!\rangle=0$ we deduce
$\langle[X^i,X^j]^{ab}\rangle=0$. This means
that the classical ground state of the model (the minimum of the bosonic
potential $\sum_{i<j}\tr[X^i,X^j]^2$) is that wherein the D-brane coordinates
commute and have simultaneous eigenvalues corresponding to definite D0-brane
positions. The full matrix model, which incorporates quantum fluctuations about
the classical ground state, thus describes smeared-out D0-brane configurations
in a spacetime with a non-commutative geometry \cite{wittenD,bfss,Tduality}.

More generally, we note that the Nicolai map $X\to W(X)$ is many-to-one, so
that general correlators of the $X$ matrices can have a
multi-valued branch cut structure. To see if this is the case, we use
the Nicolai field to write down a set of Schwinger-Dyson equations for the
matrix model. The basic identity follows from the formula for Gaussian
averages of products of even numbers of the fields $W_i^{ab}$,
\beq\new{\begin{array}{c}
\left\langle\!\!\left\langle W_{i_1}^{a_1b_1}W_{i_2}^{a_2b_2}\cdots
W_{i_{2n-1}}^{a_{2n-1}b_{2n-1}}W_{i_{2n}}^{a_{2n}b_{2n}}\right\rangle\!\!
\right\rangle=\left(\frac{g^2}N\right)^n
\left(\delta_{i_1i_2}\delta_{i_3i_4}\cdots\delta_{i_{2n-1}i_{2n}}
+\Pi^{(i)}[i_1,i_2,\dots,i_{2n}]\right)\\\times\left
(\delta_{a_1b_2}\delta_{a_2b_1}\delta
_{a_3b_4}\delta_{a_4b_3}\cdots\delta_{a_{2n}b_{2n-1}}+\Pi
^{(a,b)}[a_1,b_1;a_2,b_2;\dots;a_{2n},b_{2n}]\right)\end{array}}
\label{gausscorrs}\eeq
where $\Pi$ contains the sum of delta-functions over all permutations of
indices. The delta-functions in the indices $i_k$ come from the $SO(8)$
invariance of the reduced matrix model, while those in the indices $a_k,b_k$
arise from $U(N)$ invariance. The non-vanishing correlation functions of the
model are those which respect both of these symmetries.

As an explicit example, we set $n=2$ in (\ref{gausscorrs}) and sum over
$i_1=i_2$, $i_3=i_4$ and $a_1=b_2$, $a_2=b_1$, $a_3=b_4$, $a_4=b_3$ to get
\beq
\left\langle\frac\tr
N(\gamma_i)^{\dot\alpha}_\alpha\left[X^\alpha,X_{\dot\alpha}\right]
(\gamma^i)^{\dot\beta}_\beta\left[X^\beta,X_{\dot\beta}\right]~\frac\tr
N(\gamma_k)^{\dot\sigma}_\sigma\left[X^\sigma,X_{\dot\sigma}\right]
(\gamma^k)^{\dot\rho}_\rho\left[X^\rho,X_{\dot\rho}\right]\right\rangle=2^{10}
g^4\left(1+\frac2{N^2}\right)
\label{4ptNcorr}\eeq
In the large-$N$ limit, the expectation value of a product of invariant
operators factorizes into a product of correlators. Thus at $N=\infty$
(\ref{4ptNcorr}) becomes
\beq
\sum_{i,j}\left\langle\frac\tr N\left[X^i,X^j\right]^2\right\rangle^2=2^{10}g^4
\label{Ninf4ptcorr}\eeq
On the other hand, setting $i_1=i_3$, $i_2=i_4$ and the $a$'s and $b$'s equal
in the same way as above, we get the $N=\infty$ equation
\beq
\sum_{i,j}\left\langle\frac\tr N\left[X^i,X^j\right]^2\right\rangle=32g^2
\label{2ptcorr}\eeq
Combining (\ref{Ninf4ptcorr}) and (\ref{2ptcorr}) together we find that the
large-$N$ invariant variance of the $SO(8)$ operator $\frac\tr N[X,X]^2$ is
trivial,
\beq
\Delta^2\left(\frac\tr N[X,X]^2\right)\equiv\sum_{i,j}\left\langle\frac\tr
N\left[X^i,X^j\right]^2\right\rangle^2-\left(\sum_{i,j}\left\langle\frac\tr
N\left[X^i,X^j\right]^2\right\rangle\right)^2=0
\label{trivvariance}\eeq
Note that (\ref{trivvariance}) is a stronger statement than just the large-$N$
factorization of correlators, as it implies a non-trivial factorization over
the $SO(8)$ indices as well. This is one manifestation of the supersymmetry of
this matrix model. Similar other such identities can be derived for
higher-order correlators of the $X$-fields. In this formalism, it is also
possible to treat $n$-point connected correlation functions of the model.

The forms of the correlators above (and in particular (\ref{trivvariance}))
seem to suggest that all
non-vanishing observables in this model are analytic functions of the coupling
constant $g$ at $N=\infty$. This in turn implies that the large-$N$ limit of
the matrix model exhibits no phase transitions as one continuously varies $g$.
As the relationship with M theory dynamics is eventually obtained in the
uncompactified limit where $R\to\infty$ \cite{bfss}, the Nicolai map
demonstrates explicitly that this limit can be taken unambiguously since there
is no variation in the analytic structure of the large-$N$ solution. Moreover,
the absence of phase transitions suggests that S-duality $g\to1/g$ is
maintained in the large-$N$ limit of the matrix model above. A more precise
examination of these properties requires the inversion of the Nicolai map
(\ref{nicolaimapdef}) to get $X(W)$, which would enable one to compute
arbitrary non-supersymmetric
correlators of the matrix model. The problem in trying to construct this
inverse map is that generally $W\neq0$, corresponding to the fact that the
D-brane coordinates live in a non-commutative spacetime, so that it is not
possible to simultaneously diagonalize the $X^i$'s and find the relationship
between the eigenvalue models for the $W$ and $X$ fields. The entire
non-triviality of the matrix model lies in the correlators of invariant
combinations of the operator $X(W)$. The problem of inverting the Nicolai map
has been discussed from a perturbative point of view in \cite{lech}, where it
was also shown that this transformation is a non-polynomial functional of the
bosonic fields in the ten dimensional ${\cal N}=1$ supersymmetric Yang-Mills
theory. It would be interesting to determine this inverse map, and use it to
examine the properties of the Wilson loop correlators that we discussed earlier
in addition to the large-$N$ analyticity features of general correlators of the
matrix model. In any case, we have formally described a method in which one can
study features of the string scattering amplitudes (\ref{wilsons2}).

The results described above are only precisely valid with both the elimination
of the temporal dimension and the reduction to an $SO(8)$ spacetime symmetry
group. If we reintroduce the time dependence of the matrix
variables then the Nicolai map is determined as the non-local,
time-dependent functional ${\cal W}^i(t)=D_t^2X^i(t)-W^i(X(t))$ with $W^i$
given in (\ref{nicolaimapdef}). Then the partition function yields the
winding number of the multi-valued Nicolai map. The above results
from the reduced matrix model, such as the analyticity in the coupling constant
$g$, and hence the S-duality in the eleven dimensional compactification, show
that the simple method discussed above has the potential of providing some
insights into the structure of M theory. It would be interesting to see if the
reduced matrix model can describe other features, such as membrane interactions
\cite{berk}, of the eleven dimensional theory. It would also be interesting to
determine if the Nicolai map obtained above can be used to describe any
properties of the ten dimensional supersymmetric Yang-Mills theory itself.

\bigskip

We thank O. Lechtenfeld for comments on the manuscript and N. Mavromatos for
interesting discussions.

\end{document}